# Model Tree Based Adaption Strategy for Software Effort Estimation by Analogy

Mohammad Azzeh
Department of Software Engineering
Applied Science University
Amman, Jordan
PO BOX 133
m.y.azzeh@asu.edu.jo

*Abstract*— **Background**: Adaptation technique is a crucial task for analogy based estimation. Current adaptation techniques often use linear size or linear similarity adjustment mechanisms which are often not suitable for datasets that have complex structure with many categorical attributes. Furthermore, the use of nonlinear adaptation technique such as neural network and genetic algorithms needs many user interactions and parameters optimization for configuring them (such as network model, number of neurons, activation functions, training functions, mutation, selection, crossover,…etc.). **Aims**: In response to the abovementioned challenges, the present paper proposes a new adaptation strategy using Model Tree based attribute distance to adjust estimation by analogy and derive new estimates. Using Model Tree has an advantage to deal with categorical attributes, minimize user interaction and improve efficiency of model learning through classification. **Method**: Seven well known datasets have been used with 3-Fold cross validation to empirically validate the proposed approach. The proposed method has been investigated using various K analogies from 1 to 3. **Results**: Experimental results showed that the proposed approach produced better results when compared with those obtained by using estimation by analogy based linear size adaptation, linear similarity adaptation, 'regression towards the mean' and null adaptation. **Conclusions**: Model Tree could form a useful extension for estimation by analogy especially for complex data sets with large number of categorical attributes.

## I. INTRODUCTION

Estimation by Analogy (EBA) makes prediction for a new project by retrieving previously completed similar projects that have been encountered and remembered as historical projects [2, 7, 18, 21, 22, 23]. The effort values in the retrieved projects are reused as proposed prediction to the new project. In a few cases, particularly when the dataset is enough large and exhibit some normal characteristics, the effort of the retrieved project can be reused directly without adaptation [20]. But for others, it is common for the retrieved project to be regarded as an initial solution that should be refined to capture the differences between the new and retrieved projects [20].

Adaptation (synonymously adjustment) is a mechanism used to capture the differences between target project and most similar project(s) and then derive a new estimate [14, 20]. It is an important step in estimation by analogy as it reflects the structure of target project on the retrieved projects. Figure 1 illustrates the process of adjusted analogy based estimation. However, in literature, many adaptation techniques have been proposed to improve prediction accuracy of estimation by analogy such as using 'regression towards the mean' [11], Genetic based similarity adjustment [6], linear size adjustment [10, 14, 24], and nonlinear adjustment [16].

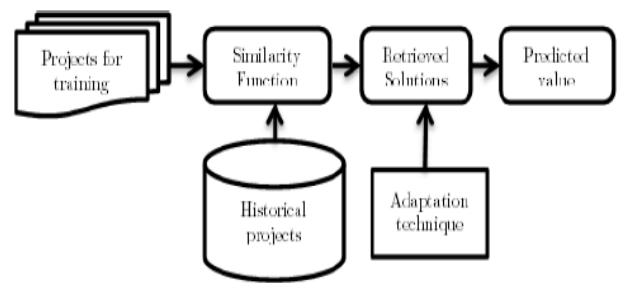

Figure 1. Process of adjusted analogy based method [16]

The majority of these adjustment mechanisms use linear adjustment such as size adjustment, similarity adjustment and productivity adjustment, which are generally restricted to size attribute and could not accept other than numeric attributes [16]. In practice, these approaches are not often efficient because software project datasets often have a complex structure and exhibit non-normal characteristics [2, 3, 16], and contain large proportion of categorical attributes [3, 8]. Moreover, the other learning based adaptation techniques such as genetic algorithm and neural networks are often challenging because they need parameter optimization and configuration setup that requires many user interactions such as decisions about: network model, number of neurons, activation functions, training functions, mutation, selection, crossover, etc. Moreover, learning and optimization through neural network and genetic algorithm takes sometimes longer time to train and may reduce performance of the model. Therefore any useful adaptation mechanism should learn from the structure of the historical dataset and should involve categorical attributes as they contain useful information to improve the accuracies of effort estimation [3, 8]. In addition to that it should minimize user interaction and reduce configuration parameters.

In response to the abovementioned reasons, the present paper proposes a new flexible adaptation technique based on Model Tree (see section 3 for more details) using attribute distance values between source historical projects and their closest analogies. In this approach, the conventional EBA procedure is first executed to produce an un-adjusted retrieval effort. Then, the differences between target project and its analogy along all attribute values are treated as inputs to Model Tree model that has been constructed based on differences of historical projects and their closest analogies to generate adjusted difference. Finally, the retrieved solution and the adjustment from Model Tree are summed up to generate the final prediction.

This paper is organized as follows: Section 2 presents related work to adaptation mechanism in analogy based estimation. Section 3 presents an overview to Model Tree. Section 4 describes the process of the proposed approach. Section 5 describes evaluation criteria used in the model validation. Datasets are described in section 6. Section 7 presents the obtained results. Sections 8 and 9 present threats to validity and conclusions respectively.

## II. RELATED WORKS

The topic of adaptation mechanism in analogy based estimation has been an active research topic in the last decade which resulted in various adaption techniques such as un-weighted mean [11, 21, 22], weighted mean [17], and median [17] as shown in Eq. (1) and (2). However, these adaptation mechanisms are directly applied to the retrieved efforts and do not capture the differences between target project and retrieved projects. Practically, the adaptation mechanisms should first capture the differences between attributes of target project and attributes of retrieved projects and then reflect these differences on the retrieved projects' effort values.

$$Effort(p_t) = \frac{1}{K} \sum_{i=1}^{K} Effort(p_i) \quad (1)$$

$$Effort(p_t) = \frac{1}{K} \sum_{i=1}^{K} w_i \times Effort(p_i) \quad (2)$$

where $K$ is the number of closest analogies, $w_i$ is the weight of project $i$. On the other hand, Walkerden and Jeffery [24] first proposed the linear size adaptation which is performed based on the linear extrapolation between size (i.e. Function Points FP) of target and source project as depicted in Eq. (3). The main restriction of this approach is that the Function Points and effort are assumed to be strongly correlated [14, 16]. In addition, the mechanism may not be applicable when size function is not Function Points. On the other hand, Mendes et al. [17] carried out several studies to check the impact of case adaptation and adaptation rules on prediction accuracy of analogy estimation. Their adaptation rules considered predefined weight of each retrieved effort according to its closeness as shown in Eq. (4). The results revealed that using adaptation rules are not significant as they did not contribute to better estimation.

$$Effort(p_t) = \frac{Effort(p_a)}{FP(p_a)} \times FP(p_t) \quad (3)$$

$$Effort(p_t) = \frac{1}{K} \sum_{i=1}^{K} \frac{1}{M} \sum_{j=1}^{M} \frac{f_{jt}}{f_{ji}} Effort(p_i) \quad (4)$$

where $M$ is the number of size features involved. $f_{jt}$ is the $j^{th}$ feature value of the target project and $f_{ji}$ is the $j^{th}$ feature value of the project $i$. $p_a$ is the closest analogy. $p_t$ is the target project. Li J. [15] proposed another adjustment approach using similarity degrees ($SM$) of all $K$ analogies as weights for effort adjustment as shown in Eq. (5)

$$Effort(pt) = \sum_{i=1}^{K} \frac{SM(pt, pi) \times Effort(pi)}{\sum_{i=1}^{K} SM(pt, pi)} \quad (5)$$

On the other hand, Jorgensen et al. [11] investigated the use of 'regression towards the mean' statistical method (see Eq. 6) to adjust the analogy estimation using the adjusted productivity of the new project and productivity of the closest analogies. This method is more suitable when the selected analogues are extreme and the estimation model inaccurate. They indicated that the adjusted estimation using 'regression towards the mean' method follows the same estimation procedure conducted by experts. This method can also be regarded as an extension of Walkerden and Jeffery's model, as it adjusts the ratio of closest productivity by adding a component of 'regression toward the mean'.

$$Effort(p_t) = FP(p_a) \times [PR(p_a) + (M - PR(p_a)) \times (1 - r)] \quad (6)$$

Where $FP(p_a)$ is function points of closest analogy. $PR(p_a)$ is the productivity of closest analogy which is measured as (Effort/FP). M is the average productivity of the similar projects. r is the historical correlation between the non-adjusted analogy based productivity and the actual productivity as a measure of the expected estimation accuracy. Chiu & Huang [6] investigated the use of Genetic Algorithms (GA) based project distance to optimize similarity between target project and its closest projects. More recently Li et al. [16] proposed alternative approach for analogy software cost estimation based on nonlinear method (Neural network). The method is suitable for complex non-uniform datasets as it has ability to learn the difference between target project and top similar projects. But the use of neural networks and genetic algorithms in adaptation mechanism need user interactions and parameter optimizations as explained in section one.

## III. MODEL TREE

The Model Tree (MT) [19, 25] as shown in Figure 2 is a special type of decision tree model developed for the task of nonlinear regression. However, the main difference between MT and regression trees is that the leaves of regression trees present numerical values, whereas the leaves of a MT have regression functions. The general model tree building methodology allows input variables to be a mixture of continuous and categorical variables. The most common approach used to implement MT is M5P algorithm which first proposed by Quinlan R. [19, 25]. M5P is regarded a powerful approach as it implements both MTs and regression trees for predicting a continuous variable [19]. The principle behind M5P is fairly simple, that is, it is constructed through a process known as binary recursive partitioning method (decision tree induction algorithm). This is an iterative process of splitting the data into partitions by minimizing the intra-sub variation in the class values down each branch, and then splitting it up further on each of the branches until the class values of all instances that reach a node vary very slightly, or only a few instances remain. The pruning and smoothing procedures are then applied respectively to the built tree to remove unwanted nodes and avoid sharp discontinuities between adjacent linear models at the leaves of the pruned tree. The full implementation of M5P algorithm is available as MATLAB code in [26].

```
if z5 = 0
  if z3 = 0
    y = 0.017929802 -2.0348163*z1 (23)
  else
    if z2 = 0
      if z4 = 0
        y = -0.0015175334 +0.97185955*z1 (10)
      else
        y = 0.012205123 (15)
    else
      if z4 = 0
        y = -0.07698552 +1.0910505*z1 (11)
      else
        y = 1.0122469 (16)
else
  y = 2.9886087 (25)
Number of rules in the tree: 6
```

Figure 2. Sample of MT [26]

The real strength of MTs, however, lies in their inherent simplicity, and the ease with which they can be interpreted by non-experts in either computing or the particular application subject. The main reason behind using MT in effort adjustment is due to its efficiency in modeling nonlinear relationship between feature difference values (as input) and effort difference values (as output) enabling us to capture the possible difference between target project and its closest projects and reflects that on the final prediction. Also it is better than traditional linear techniques in allowing for interactions and nonlinearities when numerous predictors are present. The procedure of how to implement MT for the problem of effort adjustment is explained in section 4.1.

## IV. THE ADJUSTED ANALOGY BASED ESTIMATION METHOD

The main objective of the adjustment is to capture the 'update' that transforms the effort from the retrieved projects into the target effort. When a new target project comes to be predicted, the conventional analogy based estimation method is first executed to produce an un-adjusted retrieval effort. Then, the differences between the target project's attributes and its analogy's attributes are treated as input to MT model that is being constructed based on differences of historical projects and their closest analogies as explained in subsection *A*. This will result in an adjusted difference between target project and its closest analogy. Finally, the retrieved solution and the adjustment from MT are summed up to generate the final prediction. The code of the proposed model and other comparative adaptation techniques are implemented using MATLAB. The complete procedure of prediction and adjustment is illustrated in two stages (subsections *A* and *B*).

### A. Stage I: constructing MT –based adaptation mechanism for a target project

In this stage the MT based adaptation strategy is constructed based on historical projects (training dataset) using **Jackknife** procedure as explained below.

1. Project number $i$ ($P_i$) is removed from training dataset as test project and the remaining projects are treated as historical projects.
2. The most similar project ($P_a$) to the test project is retrieved using Euclidean distance in Eqs. (7) and (8).

$$SM(P_i, P_j) = \sqrt{\sum_{k \in M} \Delta(P_i, P_j)}, \quad j = 1,2,...,n \quad 1 \quad (7)$$

where *SM* is the similarity measure. *M* is the number of predictor attributes, $P_i$ and $P_j$ are projects under investigation and:

$$\Delta(P_i, P_j) = \begin{vmatrix} (P_{ik} - P_{jk})^2 & \text{if } k\text{'s value is continuous} \\ 0 & \text{if } P_{ik} = P_{jk} \text{ (categorical)} \\ 1 & \text{if } P_{ik} \neq P_{jk} \text{ (categorical)} \end{vmatrix} \quad (8)$$

3. Compute the difference between test project ($P_i$) and its closest analogy ($P_a$) in terms of each attribute individually in addition to the difference between their effort values as shown in Eqs. (9) and (10).

$$d_k(P_i, P_a) = P_{ik} - P_{ak}; \quad k = 1,2,...,M; \quad (9)$$
$$d_e(P_i, P_a) = P_{ie} - P_{ae} \quad (10)$$

where: $d_k$ is the difference between $i^{th}$ project ($P_i$) and its closest analogy $P_a$ at $k^{th}$ attribute. $d_e$ is the difference between $i^{th}$ project ($P_i$) and its closest project $P_a$ at effort attribute.

4. Project *i*, which had been removed from the data set, is added back and steps 1 to 3 are repeated until all individual projects are evaluated. The results of this process are stored in Table 1.
5. After obtaining differences between all projects and their closest projects as depicted in Table 1, the MT based adaptation mechanism is constructed using M5P algorithm as illustrated in section 3. For this purpose we used MATLAB implementation of M5P that is developed by Jekabsons G. [26].

### B. Stage II: Prediction
1. when a new target project ($P_t$) comes to be predicted, the most similar project ($P_S$) to the target project is retrieved from the training dataset using Eq. 7.

2. The difference between them is calculated along all predictable attributes using Eq. 11 which forms input to the MT that has been built in stage I.

$$d_k(P_t, P_S) = P_{tk} \quad P_{Sk}; \quad k = 1, 2, ..., M; \quad (11)$$

3. The differences from Eq. 5 are entered as input into the constructed MT model in stage I in order to predict difference in the effort value between target project and its closest analogy ($d_e(P_t, P_S)$).

4. The predicted $d_e(P_t, P_S)$ obtained from MT model is then used to adapt effort for target project as shown in Eq. (12).

$$Effort(P_t) = Effort(P_S) + d_e(P_t, P_S) \quad (12)$$

TABLE 1 Attribute value differences between historical projects and their closest analogies in the training data set.

| Input | | | Output |
|---|---|---|---|
| $d_1(P_1, P_{a1})$ | .... | $d_M(P_1, P_{a1})$ | $d_e(P_1, P_{a1})$ |
| $d_1(P_2, P_{a2})$ | .... | $d_M(P_2, P_{a2})$ | $d_e(P_2, P_{a2})$ |
| $d_1(P_3, P_{a3})$ | .... | $d_M(P_3, P_{a3})$ | $d_e(P_3, P_{a3})$ |
| ... | .... | ... | ... |
| $d_1(P_N, P_{aN})$ | .... | $d_M(P_N, P_{aN})$ | $d_e(P_N, P_{aN})$ |

## V. EVALUATION CRITERIA

To evaluate the accuracy of the proposed estimation method, we have used common evaluation criteria in the field of software effort estimation. Mean Magnitude Relative Error (MRE) computes the average of absolute percentage of error between actual and predicted effort for each project.

$$MMRE = |P|^{-1} \sum_{p_i \in P} \frac{|Effort(p_i) - \overline{Effort(p_i)}|}{Effort(p_i)} \quad (13)$$

Where $Effort(p_i)$ and $\overline{Effort(p_i)}$ are the actual value and predicted values of project $p_i$. MdMRE calculate the mean and median of MRE over all reference projects [13]. We also used Boxplot of absolute residuals and Wilcoxon signed rank test to investigate the statistical significance of all the results, setting the confidence limit at 0.05. PRED ($\ell$) is used as a complementary criterion to count the percentage of estimates that fall within less than $\ell$ of the actual values. The common used value for $\ell$ is 25%.

$$PRED(\ell) = \frac{\lambda}{N} \times 100 \quad (14)$$

Where $\lambda$ is the number of projects that have $MRE \leq \ell\%$, and $N$ is the number of all observations.

## VI. DATASETS

Seven datasets have been used for the purpose of model evaluation. These datasets come from different sources: ISBSG (release 10, 2007) [9], Desharnais [5], Kemerer [12], Albrecht [1], COCOMO'81 [4], Maxwell and China datasets [5]. The descriptive statistics of the datasets are given in Table 2. All datasets and their attributes information are available at PROMISE website except ISBSG which is not available for public use due to non-disclosure agreement.

TABLE 2 Descriptive statistics of the datasets

| Dataset | Cases | Categorical Features | Numerical Features | Effort mean | Effort Std. |
|---|---|---|---|---|---|
| ISBSG | 500 | 1 | 8 | 14668 | 11727 |
| Desharnais | 77 | 1 | 10 | 4834 | 4188 |
| COCOMO'81 | 63 | 2 | 15 | 406.4 | 657 |
| Kemerer | 15 | 2 | 4 | 219.25 | 263 |
| Albrecht | 24 | 1 | 6 | 21875 | 28417 |
| Maxwell | 62 | 22 | 3 | 8223 | 10500 |
| China | 499 | 0 | 18 | 3921 | 6481 |

## VII. RESULTS

### A. Evaluation Procedure

Three fold cross validation has been used to validate the accuracies of the proposed method, similarly to Mendes et al. [17] and Li et al. [16]. The data set is randomly divided into three equally sized subsets. At each time, two of the three subsets are used as the training set in order to construct the adjusted estimating model. The remaining subset is used as testing set exclusively for evaluating model prediction. The evaluation procedure is repeated 3 times in that every candidate set is held out once for testing and the prediction model is trained on the remaining observations, then their MRE values and residuals are evaluated. Finally, the evaluation results are aggregated across all validation sets. To investigate the impact of number of analogy (K) we vary K from 1 to 3, similarly to Kirsopp et al. [14], Li et al. [16] and Mendes et al. [17], in that we used un-weighted mean to aggregate adapted effort in case of K=2 and 3. All numeric attributes have been normalized using min-max normalization schema in order to have the same influence. Also all datasets were pre-processed to avoid missing values by removing the entire records with missing values.

The proposed MT based adjusted estimation by analogy method with best variants has also been compared to conventional estimation by analogy (EBA) [22], and other well-known adjusted analogy based estimation methods such as: Linear size adjustment (L-EBA, see Eq. 3) [24], Linear similarity adjustment (S-EBA, see Eq. 5) [15] and regression towards the mean (R-EBA, see Eq. 6) [11]. It is important to know that the Euclidian distance in Eq. (7) was the common similarity measure for all EA based adaptation techniques. Unfortunately, because we do not have precise details as to how genetic [6] and neural networks [16] parameter were optimized, it was difficult to validate MT-EBA against nonlinear adjustment mechanisms [6, 16]. For this reason, we left this part of research as future work.

TABLE 3 Prediction accuracy of MT-ABE over different datasets using different analogy numbers

| Dataset | MMRE% | MdMRE% | PRED% |
|---|---|---|---|
| K=1 | | | |
| ISBSG | 24.2 | 22.2 | 61.8 |
| Desharnais | 42.7 | 25.6 | 50.7 |
| COCOMO | 26.5 | 24.7 | 51.7 |
| Kemerer | 57.6 | 28.7 | 33.3 |
| Albrecht | 39.1 | 32.8 | 37.5 |
| Maxwell | 69.8 | 18.6 | 56.5 |
| China | 45.7 | 11.9 | 64.7 |
| K=2 | | | |
| ISBSG | 23.6 | 21.8 | 68.3 |
| Desharnais | 36.2 | 25.6 | 50.7 |
| COCOMO | 21.7 | 21.9 | 60 |
| Kemerer | 36.5 | 26.1 | 46.7 |
| Albrecht | 32.3 | 19.9 | 58.3 |
| Maxwell | 76.5 | 49 | 54.8 |
| China | 40.1 | 11.4 | 65.3 |
| K=3 | | | |
| ISBSG | 20.1 | 19.5 | 62 |
| Desharnais | 26.1 | 12 | 72.7 |
| COCOMO | 23.3 | 21.6 | 53.3 |
| Kemerer | 45.2 | 29.9 | 33.3 |
| Albrecht | 34.9 | 26.6 | 45.8 |
| Maxwell | 73.2 | 47 | 50 |
| China | 34.9 | 10.9 | 67.1 |

*B. Prediction Accuray of MT-EBA*

This section presents empirical validation of MT-EBA estimation model over the employed data sets. The results shown in Table 3 summarize the relative accuracy of the MT-EBA with different analogy numbers using the MMRE, MdMRE and PRED values for all data sets. In brief, the obtained results for all data sets (except for Maxwell dataset) are promising as being more predictive especially in terms of MMRE. The notable results from this table are for Desharnais, ISBSG and NASA93 data sets where MT-EBA obtained good estimates with MMRE and MdMRE less than or around 25%. Although the performance figures on Maxwell data set was poor, it is still considered promising if we compared it to other published results on Maxwell data set such as those obtained by Li et al. [16].

As can be seen from Table 3, using two and three analogies performs better for most of the employed datasets. MT-EBA (K=2) performs better in terms of MMRE and PRED for three data sets including Albrecht, Kemerer and NASA93, indicating the potential improvements of MT-EBA with few number of analogies for relatively small datasets. However, while this is not surprising, it is important to know that the choice of adaptation technique and number of analogies do matter. On the other hand, it is widely acknowledged that for large data sets the choice of one analogy is quite sufficient because it is more likely to find similar projects within large case-base [3, 8]. This was not the case for ISBSG and China datasets where the use of MT-AEBA with K=3 was notably superior to K=1 and K=2 in terms of MMRE. In general, we can conclude that all results for MT-EBA with K=1, 2 and 3 were good, and corroborate that if the adjustment is learnt from historical projects we can obtain accurate estimates than using non-adjusted solutions.

TABLE 4 MMRE comparisons between MT-EBA & other methods

| Dataset | MT-EBA | EBA | R-EBA | L-EBA | S-EBA |
|---|---|---|---|---|---|
| ISBSG | **20.1** | 74.3 | 55.4 | 78.5 | 80.1 |
| Desharnais | **26.14** | 65.9 | 57.9 | 56.7 | 71.2 |
| COCOMO | **21.7** | 66.4 | 41.5 | 58.3 | 47.5 |
| Kemerer | **36.5** | 54.3 | 47.2 | 60.2 | 77.9 |
| Albrecht | **32.3** | 79.6 | 50.0 | 68.9 | 61.8 |
| Maxwell | 69.8 | 133.7 | **55.6** | 92.8 | 61.2 |
| China | **34.9** | 59.6 | 70.9 | 76.5 | 98.8 |

*C. Comparison MT-EBA to Other Adaptation Based Estimation by Analogy Techniques.*

The best variants from Table 3 are selected to compare MT-EBA with best variants of other well-known adaptation techniques based estimation by analogy: R-EBA, S-EBA, L-EBA and conventional EBA. We should note that analogously to MT-EBA we vary K from 1 to 3 for EBA, S-EBA and L-EBA and the best variants have been selected for comparison, while R-EBA used only the closest analogy as suggested in its original approach [11]. The results shown in Tables 4, 5 and 6 revealed that the predictions generated by MT-EBA are more accurate than other methods especially in terms of MMRE and PRED with one exception for Maxwell dataset. The main reason for that may be related to the strong correlation between predicted productivity and project size in this dataset. The results also surprisingly reveal that S-EBA and L-EBA produced worst accuracy than EBA especially for large datasets (ISBSG and China) as these dataset have high probability to contain similar analogies and that kind of adjustments rely intensively on the correlation between size attribute and effort.

The results also show that R-EBA produced better results than L-EBA, S-EBA and EBA, which demonstrates that the productivity adjustment is more accurate and reliable than similarity and size adjustment. The obtained superior results for MT-EBA demonstrate the importance of involving categorical attributes in adaptation mechanism. Inevitably, we should not overlook some important factors that contribute to these superior results including choosing number of analogies and the procedure of constructing MT. The results of Wilcoxon signed rank test (Table 7) shows unsurprisingly that predictions based on MT-EBA method present statistically significant accurate estimates, measured using absolute residuals. Interestingly, using MT adaptation mechanism presents significantly better results than using EBA, L-EBA and S-EBA for all datasets except Kemerer dataset. Also MT-EBA generates statistically significant predictions than R-EBA for four datasets out of seven. These results suggest that the predictions generated by MT-EBA are different than those generated by other analogy estimation based adjustment mechanisms.

TABLE 5 MdMRE comparisons between MT-EBA & other methods

| Dataset | MT-EBA | EBA | R-EBA | L-EBA | S-EBA |
|---|---|---|---|---|---|
| ISBSG | **19.5** | 41.5 | 42.3 | 43.8 | 50.7 |
| Desharnais | **12.0** | 45.7 | 48.2 | 42.3 | 41.1 |
| COCOMO | 21.9 | 49.8 | 35.4 | **19.7** | 33.0 |
| Kemerer | 26.1 | **25.1** | 37.1 | 38.8 | 58.0 |
| Albrecht | **19.9** | 61.71 | 26.2 | 42.7 | 48.0 |
| Maxwell | **18.6** | 75.67 | 54.3 | 56.1 | 45.0 |
| China | **10.9** | 39.13 | 50.0 | 38.7 | 43.2 |

TABLE 6 PRED comparisons between MT-EBA & other methods

| Dataset | MT-EBA | EBA | R-EBA | L-EBA | S-EBA |
|---|---|---|---|---|---|
| ISBSG | **62.0** | 35.6 | 31.1 | 34.3 | 28.7 |
| Desharnais | **72.7** | 23.4 | 29.8 | 35.0 | 26.0 |
| COCOMO | **60.0** | 31.7 | 38.3 | 55.0 | 43.7 |
| Kemerer | **46.7** | 46.7 | 33.7 | 26.7 | 13.3 |
| Albrecht | **58.3** | 16.7 | 45.8 | 37.5 | 25.0 |
| Maxwell | **56.5** | 11.1 | 20.9 | 27.4 | 21.0 |
| China | **67.1** | 29.3 | 25.9 | 48.5 | 30.9 |

The obtained results are also confirmed by Boxplot of absolute residuals shown in Figures 3 to 9, which reveal that MT-EBA generated generally better predictions than other adaptation methods. The Boxplots suggest that in general the absolute residuals are skewed towards the minimum value as the box of MT-EBA overlays the lower tail. The box length and median of MT-EBA are smaller than the box length and median of other methods which indicate that at least half of the predictions of MT-EBA are more accurate than other methods. It can be observed also that EBA and L-EBA and S-EBA produced many outlying estimates and again confirm that the greatest problems were encountered when the projects were at the extreme, or maximum, for the range of values. Obviously, further investigation would be useful, but the finding is intuitively reasonable that the more representative a target project is, the better the predictions are.

TABLE 7 Wilcoxon sum rank test between MT-EBA & other methods over all data sets.

| | EBA | L_EBA | S-EBA | R-EBA |
|---|---|---|---|---|
| **ISBSG** | -7.1 [a] | -7.6 [a] | -11.7 [a] | -9.33 [a] |
| **Desharnais** | -5.8 [a] | -4.7 [a] | -5.9 [a] | -5.22 [a] |
| **COCOMO** | -2.2 [b] | -0.8 | -2.0 [b] | -2.1 [b] |
| **Kemerer** | -0.17 | -1.3 | -1.94 [b] | -0.37 |
| **Albrecht** | -3.9 [a] | -2.9 [a] | -2.44 [a] | -1.8 |
| **Maxwell** | -4.8 [a] | -2.2 [b] | -1.6 | -1.6 |
| **China** | 5.4 [a] | 5.18 [b] | -2.31 [b] | -1.97 [b] |
| [a]: Significant at 1%, [b]: significant at 5% | | | | |

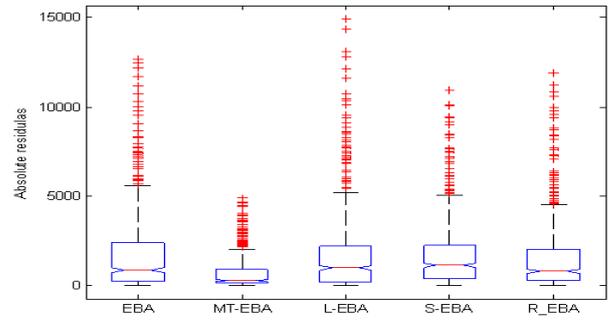

Figure 2. Boxplot of absolute residuals for ISBSG

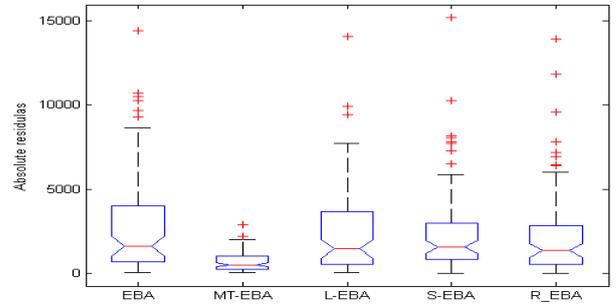

Figure 3. Boxplot of absolute residuals for Desharnais

## VIII. THREATS TO VALIDITY

This section presents the comments on the validities of our study based on the internal and external threats to validity. In our opinion the greatest threats are to the internal validity of this study; i.e. the degree to which conclusions can be drawn with regard to the better parameter setup for Analogy-based effort prediction. One possible threat to internal validity is the chosen of similarity function, given that different similarity functions yield different predictions as they retrieved different analogies. In our study we preferred to use the common similarity function (Euclidean distance) as it has been widely used in different implantations of EA. This may portrait other similarity functions as non-effective measures, however, this is not true and the efficiency of those measures has been confirmed in previous studies.

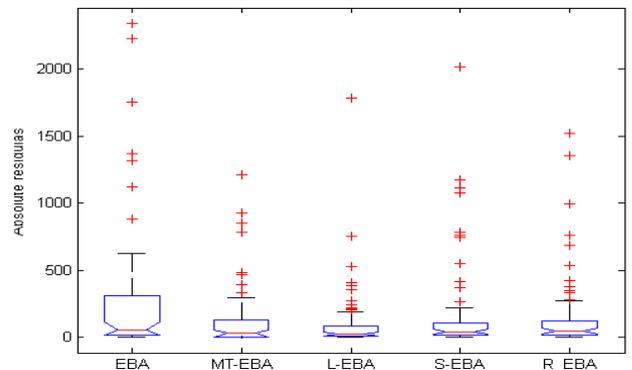

Figure 4. Boxplot of absolute residuals for COCOMO

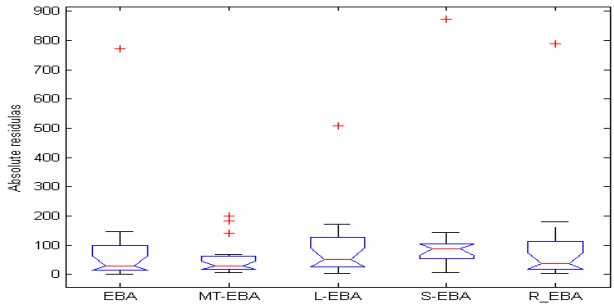
Figure 5. Boxplot of absolute residuals for Kemerer

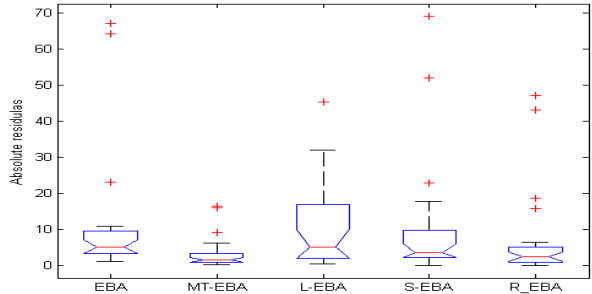
Figure 6. Boxplot of absolute residuals for Albrecht

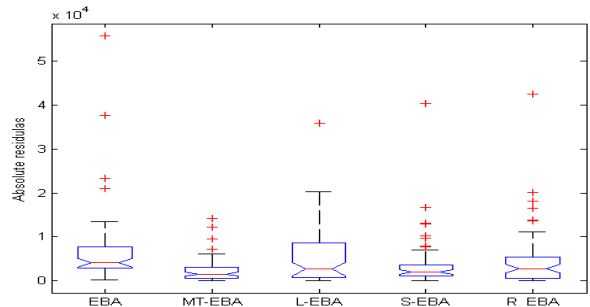
Figure 7. Boxplot of absolute residuals for Maxwell

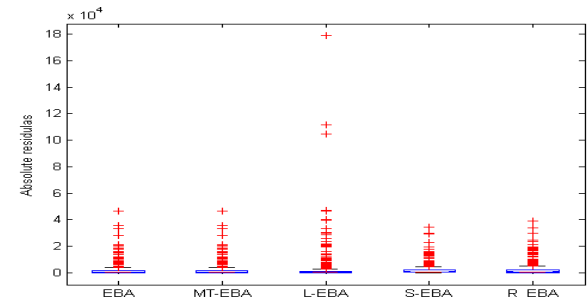
Figure 8. Boxplot of absolute residuals for China

Despite special emphasis was placed on the effectiveness of the number of analogies, complete certainty with regard to this issue was impossible and we had to rely on limited number of analogies from 1 to 3. We do not consider that choice was a problem since this study was motivated with the finding from previous studies [14, 17]. On the other hand, for better use of R-EBA it was recommended to divide the dataset into homogenous groups in order to increase prediction accuracy [11]. This recommendation has not been performed because of the complexity to find the feature that reasonably divides the dataset into homogenous groups, so we left this point for further investigation. Finally, In order to make apple-to-apple comparisons between different adaptation techniques we preferred to use 3-Fold cross validation strategy, though some authors favoured Jack Knifing validation. The principal reason is that the 3-Fold cross validation is more appropriate to compare between different prediction models as recommended by Kirsopp et al. [14] and Mendes et al. [17]. With regard to external validity, i.e. the ability to generalize the obtained findings of our comparative studies, we consider that some datasets are very old to be used in software cost estimation because they represent different software development approaches and technologies. The reason to this is that these datasets are free, and still widely used for comparison purposes.

## IX. DISCUSSION AND CONCLUSIONS

This paper proposed a new automatable adjustment mechanism based Model Tree which does not need user interactions as in nonlinear adjustment mechanism. This yields the advantage of avoiding explicit knowledge elicitation. It can also be applied without prior change or tuning for different data sets containing different attributes. This paper has presented a development of the original formulation of the nonlinear adjustment mechanism that is more generally applicable to EBA. The proposed method can be now applied in the absence of feature selection and where features are negatively correlated to effort. It also adds a mechanism - supported by Model Tree process- for choosing which features are appropriate for adaptation for each target project (i.e. regression equations derived by MT are different for each target project). Through evaluation of seven datasets we demonstrated that effort estimates can be significantly improved through Model Tree adjustment. Six of the seven data sets showed in general statistically significant improvements in prediction accuracy when using MT-EBA whilst the smallest data set (Kemerer) showed no significant difference so it seems that any adjustment mechanism generally improves the accuracy of the predictions for this dataset. Further, from the results, MT-EBA presents number of interesting advantages. First, MT-EBA remains viable when using too many categorical features (e. g. the Maxwell data set). Second, MT-EBA remains accurate for small data sets (e. g. the Albrecht (24 projects), Kemerer (15 projects)), and for large data sets as well (e.g. ISBSG and China). Third, MT-EBA remains accurate where the number of features is limited (e. g. the Kemerer data set).

Another interesting point is that, for all of the data sets under consideration, all attributes have been involved (without pruning) in developing Model Tree adjustment mechanism which indicates the effectiveness of MT-EBA. But we still need to investigate the effect of feature subset selection on the accuracy of MT-EBA. However, this analysis has also permits an empirical evaluation of a number of points relating to the most effective use of MT-EBA method: First, the use of MT can cause a problem when number of training projects is very low. Even so, the fact remains that estimate from very small data sets should be treated with caution. The second point is the selection of optimum number of analogies for MT-EBA to search for. The answer to the question appear to be subjective in that 'three Analogies' is the most commonly accurate estimation method for large data sets (e.g. ISBSG and China), being selected for 3 out of the 7 datasets. 'Two Analogies' was the most accurate 3 times, for small and medium data sets (e.g. Kemerer and NASA93), whilst 'One Analogies' is most accurate for the data set with too many categorical features. The main assumptions made in previous studies [3,8,14] was that the selection of just one analogy would be suitable choice for large data sets while a smaller data set would favour the selection of more analogies. This however, has not been ascertained in the results, with for example, the two largest data sets, ISBSG and China finding respectively three analogies to be the optimum number to search for. Even though the selection of 'One Analogy' seems to be superior for Maxwell dataset which is considered somehow medium in size, it must be remembered that for many of the data sets the use of three different analogies methods returned remarkably predictive accuracy levels. Perhaps the only conclusion that can be drawn on this point is that, the larger the data set, the more consistent the results are likely to be. This point needs further investigations and will be looked at again from a different view point in the future works, where individual data sets will be examined to see if accuracy improves as more analogies are utilized. So another study is required to investigate the impact of using more analogies than three on the accuracy sensitivity of MT-EBA. Further empirical investigation is also necessary to ensure the validity of proposed adjustment mechanism on other datasets and in the presence of feature subset selection. Future extension of the proposed model is planned to compare our proposed approach with nonlinear adjustment mechanism Such as genetic based similarity and neural network adjustment.

ACKNOWLEDGMENT

The author is grateful to the Applied Science University, Amman, Jordan, for the financial support granted to cover the publication fee of this research article.